\documentclass[runningheads]{llncs}
\usepackage[T1]{fontenc}
\usepackage{float}

\usepackage{graphicx}
\usepackage{setspace}
\usepackage{amsmath}

\begin{document}

\title{Bloom Filter Look-Up Tables for Private and Secure Distributed Databases in Web3\thanks{Partially supported by the Israeli Science Foundation (Grant No. 465/22), the Rita Altura Trust Chair in Computer Science, and the Frankel Center for Computer Science. An initial version appeared in DBSec 2025 and is updated and corrected in this revised version.}\\
{\large\rm (Revised Version)}}

\titlerunning{BFLUT Based Private and Secure Distributed DB for Web3}

\author{Shlomi Dolev
\and 
Ehud Gudes
\and Daniel Shlomo
}
\authorrunning{S. Dolev, E. Gudes, D. Shlomo}
\institute{Ben-Gurion University of the Negev, Beer-Sheva, Israel\\
\email{dolev@cs.bgu.ac.il, ehud@cs.bgu.ac.il, danshl@post.bgu.ac.il}}

\maketitle

\nocite{
Benet2014IPFS,
gennaro1999distributed,
zheng2017blockchain,
OrbitDBDocs,
crdtTech,
kamara2010cryptographic,
Dolev2022BFLUT,
zhang2019security,
heilman2015eclipse,
IPNSDocs,
androutsellis2004p2p}

\begin{abstract}
The rapid growth of decentralized systems in the Web3 ecosystem has introduced numerous challenges, particularly in ensuring data security, privacy, and scalability \cite{zheng2017blockchain,zhang2019security}. These systems rely heavily on distributed architectures, requiring robust mechanisms to manage data and interactions among participants securely. 

One critical aspect of decentralized systems is key management, which is essential for encrypting files, securing database segments, and enabling private transactions. However, securely managing cryptographic keys in a distributed environment poses significant risks, especially when nodes in the network can be compromised \cite{heilman2015eclipse}. 

This research proposes a decentralized database scheme specifically designed for secure and private key management. Our approach ensures that cryptographic keys are not stored explicitly at any location, preventing their discovery even if an attacker gains control of multiple nodes. Instead of traditional storage, keys are encoded and distributed using the BFLUT (Bloom Filter for Private Look-Up Tables) algorithm \cite{Dolev2022BFLUT}, which enables secure retrieval without direct exposure. 

The system leverages OrbitDB \cite{OrbitDBDocs}, IPFS \cite{Benet2014IPFS}, and IPNS \cite{IPNSDocs} for decentralized data management, providing robust support for consistency, scalability, and simultaneous updates. By combining these technologies, our scheme enhances both security and privacy while maintaining high performance and reliability.

Our findings demonstrate the system's capability to securely manage keys, prevent unauthorized access, and ensure privacy, making it a foundational solution for Web3 applications requiring decentralized security.
\end{abstract}

\section{Introduction}

The rise of decentralized systems in the Web3 environment introduces unique challenges in securely and efficiently managing critical data \cite{zheng2017blockchain}. While Web3 provides transparency, self-sovereignty, and decentralized data sharing without reliance on central intermediaries, its peer-to-peer (P2P) architecture requires robust mechanisms to ensure privacy, security, and conflict-free updates \cite{androutsellis2004p2p}. These challenges are amplified by the need to maintain availability and consistency in a distributed network where multiple nodes participate in data storage and dissemination.

One of the key considerations in decentralized systems is how to balance privacy and accessibility. While data must remain accessible to authorized users, it is essential to prevent unauthorized access and ensure that even if some nodes are compromised, sensitive information remains secure. Traditional approaches often rely on explicit data storage, which can expose sensitive information to potential attacks or breaches.

In this research, we propose a novel decentralized database scheme designed to securely store and retrieve keys without directly storing them. Unlike conventional systems, where sensitive data or keys are explicitly stored, our scheme encodes information in a distributed manner across the network by using a combination of advanced technologies. This ensures that even if certain nodes are compromised, the encoded information cannot be reconstructed or exploited by attackers.

Our system leverages the following components:

\begin{description}
    \item [\textbf{OrbitDB and CRDTs:}]
        OrbitDB \cite{OrbitDBDocs} is a distributed database built on IPFS \cite{Benet2014IPFS}, enabling decentralized storage without centralized intermediaries. It supports various database types, such as logs, key-value stores, and document-based systems, making it adaptable for decentralized environments.  

        A key feature is its use of CRDTs (Conflict-Free Replicated Data Types) \cite{crdtTech}, which ensure data consistency across nodes by allowing concurrent updates without conflicts.
    
   \item[ \textbf{IPNS (InterPlanetary Name System):}]
        IPNS \cite{IPNSDocs} is a protocol for persistent addressing in decentralized networks, built on top of IPFS \cite{Benet2014IPFS}.\\ In IPFS, files are addressed by their content identifier (CID), which changes whenever the content is modified. This dynamic nature complicates access to frequently updated files, as older references become obsolete.
        
        IPNS solves this by associating a fixed, cryptographically signed name with the file’s changing CID, allowing seamless access to the latest version. In our system, this ensures reliable data retrieval, even as files are updated, simplifying management in distributed environments with frequent modifications.

    \item [\textbf{BFLUT (Bloom Filter for Private Look-Up Tables):}]
        At the core of our solution is the BFLUT algorithm \cite{Dolev2022BFLUT}, which encodes keys into secure, distributed representations. This approach avoids explicitly storing cryptographic keys, ensuring that they cannot be directly exposed or reconstructed, even if an attacker gains access to certain nodes in the network.  .\\
        BFLUT is built upon the Bloom Filter technique, a probabilistic data structure that efficiently verifies whether an element is present in a set without storing the element itself. This enables BFLUT to securely encode keys by activating specific bits in the Bloom Filter, determined by hash functions applied to the key's components. 
        \begin{spacing}{1.5}
        \end{spacing}
        This mechanism is vital for our decentralized database, as it allows us to manage keys securely and privately, enabling efficient key retrieval while maintaining robust protections against unauthorized access (see Section ~\ref{sec:BFLUTExplain}).
\end{description}

This work contributes to the design and analysis of a secure decentralized database, specifically focusing on efficient key encoding and retrieval. By addressing privacy and scalability concerns, our approach lays the foundation for secure, reliable, and efficient Web3-based applications. \\

The rest of this paper is structured as follows:
Section 2 discusses related work, Section 3 highlights the key advantages of our approach, Section 4 outlines the methodology, and Section 5 presents the analysis of our scheme. Finally, discussions and conclusions appear in Section \ref{s:dc}. Many details are omitted from this extended abstract.

\section{Related Work}

Key management and secure data distribution are critical aspects of decentralized systems. Two significant studies provide insights into current approaches and their limitations, particularly regarding \textbf{secret sharing schemes} and \textbf{threshold cryptography}.

\subsubsection{Secure and Effective Key Management Using Secret Sharing Schemes in Cloud Computing}
This study \cite{kamara2010cryptographic} explores the use of secret sharing schemes (SSS) to enhance key management in cloud environments. The approach involves dividing a secret into multiple shares, distributed across different nodes. A predefined threshold of shares is required to reconstruct the key, thereby reducing the risk of single points of failure and improving overall security in cloud setups.

However, there are notable limitations:
\begin{enumerate}
    \item \textbf{Centralized Dependency:} The reliance on cloud providers for the distribution and reconstruction of shares means that data still resides within centralized systems, which is contrary to the decentralized philosophy of Web3.
    \item \textbf{Reconstruction Delays:} Share reconstruction in dynamic or unreliable environments can introduce significant delays.
\end{enumerate}

\subsubsection{Secure Key Management in Distributed Systems: Challenges and Solutions}
This study \cite{gennaro1999distributed} builds on the principles of secret sharing by incorporating threshold cryptography for robust key management in distributed networks. Unlike traditional approaches, this method enables operations directly on the shares without requiring full reconstruction. This improves fault tolerance, as the system can function even if certain nodes are unavailable, and enhances resilience to failures.

While the approach provides theoretical improvements:
\begin{enumerate}
    \item \textbf{Computational Overhead:} Performing operations on shares without full reconstruction increases computational complexity, especially in large-scale systems with numerous nodes.
    \item \textbf{Threshold-Based Complexity:} Systems requiring frequent key updates or decentralized storage face challenges in maintaining efficient operations, as the threshold model adds complexity to key updates.
\end{enumerate}

\subsubsection{Relevance to Our Work}

The literature review explores decentralized systems like IPFS \cite{Benet2014IPFS} and OrbitDB \cite{OrbitDBDocs}, which provide reliable, self-sovereign data storage in Web3 environments. It also examines conflict management methods, particularly CRDTs \cite{crdtTech}, which ensure consistency in distributed updates, and Bloom Filter-based techniques, such as BFLUT \cite{Dolev2022BFLUT}, for private data retrieval.  

While each of these tools operates independently, this research combines these methodologies. Existing solutions mainly focus on secure key management in controlled or semi-controlled environments like hybrid cloud systems. In contrast, our approach is designed for fully decentralized Web3 environments, offering:

\begin{enumerate}
    \item \textbf{Decentralized File Representation:} Utilizing BFLUT for bitwise modifications, reducing false positives in key retrieval on IPFS.
    \item \textbf{True Decentralization:} Eliminating reliance on a single entity, enhancing security and privacy in distributed storage.
\end{enumerate}

\section{Key Advantages of the Proposed System}
\subsubsection{High Privacy}
The system leverages the BFLUT (Bloom Filter for Private Look-Up Tables) algorithm to encrypt and securely store data. This algorithm encodes data so that it can only be accessed through a controlled retrieval mechanism. This design prevents direct exposure of sensitive information, even in cases of system breaches.

\subsubsection{Consistency and Concurrent Updates}
Using CRDTs (Conflict-Free Replicated Data Types), the system supports concurrent updates while maintaining absolute consistency across all nodes. This decentralized update mechanism eliminates the need for central synchronization, ensuring that data remains accurate and up-to-date across the entire network.

\subsubsection{Efficient and Secure Data Distribution}
The system distributes data intelligently across different nodes in the network, where each fragment of the database is stored on a separate node. This approach ensures balanced workload distribution, high availability, and fast access to required data without relying on a single point of failure. The optimized distribution mechanism also facilitates efficient file management and retrieval within the network.

\subsubsection{Security and Resilience in a Decentralized Environment}
The system is designed so that each node stores only data fragments. In contrast to traditional databases, where all the information is stored in a single geographic location and could be lost entirely due to accidental deletion or failure, our approach distributes the data across multiple nodes. This ensures that such a catastrophic loss does not occur.\\
In the event of a node failure, the data remains accessible through other nodes, maintaining availability and integrity. Moreover, even if all the files associated with a specific bit of a key were to be deleted, the system can still proceed with key extraction by treating all possible values for that bit as valid. Advanced encryption methods prevent unauthorized access, while the distributed nature ensures that even attacks targeting multiple nodes cannot compromise the data's availability or integrity.

The combination of these technologies enables the creation of a distributed, secure, and efficient database tailored to the complex requirements of the Web3 environment. Our solution addresses the intersection of privacy, availability, and update management, representing a significant advancement in distributed data management.

\section{Methodology}

Before delving into the implementation details, we first examine the general structure and the BFLUT data structure, which forms the backbone of the proposed system. BFLUT is the core mechanism for secure and efficient data encoding and storage in a decentralized environment, ensuring privacy and seamless access.

\subsection{Introduction: System Purpose and General Structure}  

The proposed system provides a secure, decentralized database within the Web3 environment. Data entries are stored as files on distributed nodes, with their references managed using a key-value structure in \textbf{OrbitDB}, a distributed database optimized for decentralized storage. Each key-value pair is structured as follows: \\ 
\noindent\textbf{Key:} A unique identifier generated using a hash function.\\  
\noindent\textbf{Value:} A reference to a file stored on a specific node in the network.  

Since files in the system are addressed by theirContent Identifier (CID)—a hash of their content—any modification results in a new CID, invalidating previous references. To maintain persistent access, the system employs InterPlanetary Name System (IPNS), which associates a fixed, cryptographically signed name with the latest CID. This ensures that even when a file is updated, its reference remains valid and accessible.  

Each file is stored on a limited number of nodes to enhance privacy, scalability, and fault tolerance, reducing reliance on any single entity. The system ensures reliable and secure decentralized data storage and retrieval by integrating OrbitDB for efficient key-value management and IPNS for dynamic file referencing.

\subsection{BFLUT: A Secure Encoding Mechanism} \label{sec:BFLUTExplain}
The BFLUT (Bloom Filter for Private Look-Up Tables) algorithm is a secure and compact encoding mechanism designed to encode and store data while ensuring privacy. It leverages the principles of Bloom Filters, which allow the system to verify the existence of data without revealing or storing the data itself. This technique ensures that sensitive information remains protected even in decentralized environments.\\
Encoding involves activating specific bits in a Bloom Filter table based on hash functions. These bits are strategically chosen to represent the data in a secure and distributed manner, avoiding the direct storage of raw data.
        \begin{spacing}{1.5}
        \end{spacing}
\textit{{Example: Storing and Searching Data}}. Consider a user named ``John Smith'' associated with the value `0110'. The encoding and storage process involves computing a hash of ``John Smith'' with the prefix `0' and activating the corresponding bit in the Bloom Filter table, represented as $H(JohnSmith0)$. Subsequently, additional bits are activated for the following prefixes: \\
$H(JohnSmith01)$,  $H(JohnSmith011)$, and $H(JohnSmith0110)$. 

This process ensures that all prefixes of the value are securely encoded in the system. By activating these bits in the Bloom Filter table, the system creates a distributed representation of the data that can be securely retrieved. When the system stores these activated bits, they are effectively linked to the user's name and associated prefixes. This linkage allows the system to verify the existence of data while maintaining its encoded form, ensuring privacy and security.
\begin{spacing}{1.5}
\end{spacing}
To retrieve the value associated with the key ``John Smith,'' the system follows this process:
\begin{enumerate}
    \item Check the bits at $H(JohnSmith0)$ and $H(JohnSmith1)$. If none of them are active, it can be concluded that the key does not exist in the system.\\
    \item If a bit is active for one of the prefixes (e.g., $H(JohnSmith0)$), continue checking the subsequent prefixes: $H(JohnSmith00)$ or $H(JohnSmith01)$.\\
    \item Repeat this process, appending additional prefixes, until the desired value length is reached (assuming the length is known).
\end{enumerate}

\begin{figure}[H]
    \centering
    \includegraphics[width=1.0\textwidth]{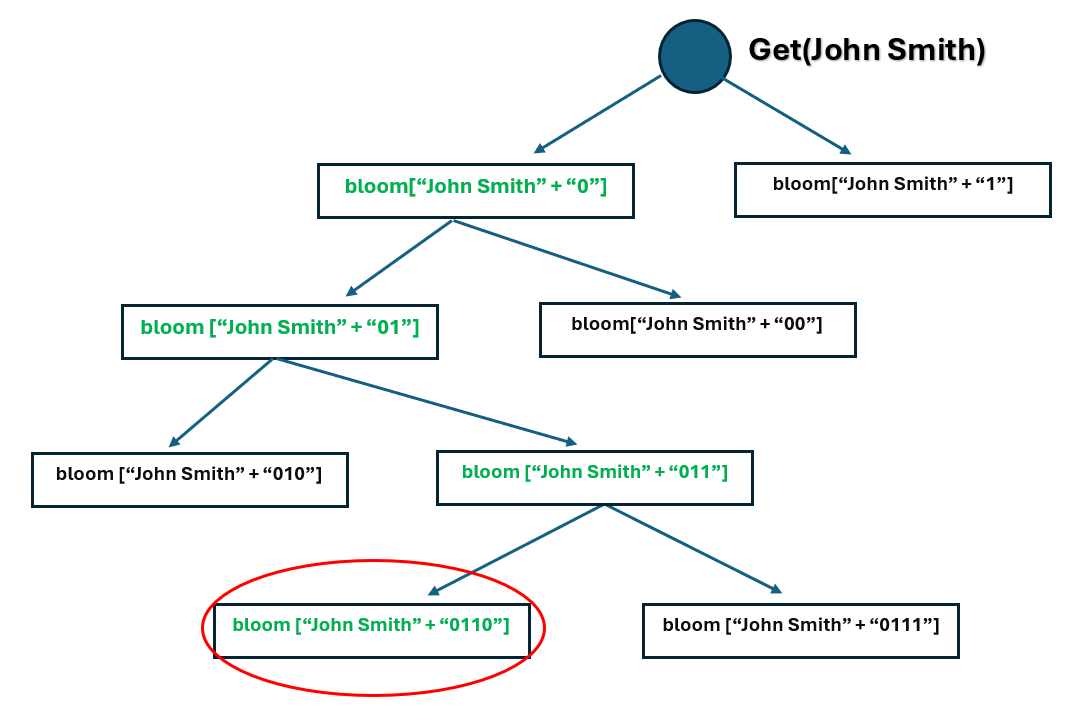}
    \caption{Get value from BFLUT example from \cite{Dolev2022BFLUT} }
    \label{fig:GetBFUT}
\end{figure}

BFLUT inherently allows for the possibility of false positives, where certain bits may appear active even if the corresponding value does not exist. However, if no solution is found, it can be conclusively determined that the key does not exist in the system. This will be elaborated in the sequel (Ref. Fig \ref{fig:GetBFUT})

\subsection{System Implementation}

Having understood the system's structure and the role of BFLUT, we now delve into the details of the system's implementation, including its initialization, data insertion, and retrieval processes.
\subsubsection{Introduction: A System for Key Management Based on Username and Password}

Let us consider a system designed to securely store and retrieve keys based on a username and password, for example, ``John Smith'' with the password ``password''. In this system, the username and password are not stored in their raw form; rather, they are used as inputs to derive a unique key (KEY). This key is chosen from a predefined range of values (in this example, binary values \(0\) or \(1\)) and is managed in a distributed and secure manner across the network.

When the correct username and password are provided, the system computes a mapping that yields the corresponding key (KEY). This derived key is then encrypted and stored across multiple nodes, ensuring that even if part of the network is compromised, the sensitive credentials are not exposed. The raw data (i.e., the username and password) is never stored directly, so the security of the system is maintained by safeguarding only the encrypted key.
 
\subsubsection{Step 1: System Initialization and File Allocation} 
\begin{figure}[h!]
    \centering
    \includegraphics[width=1.0\textwidth]{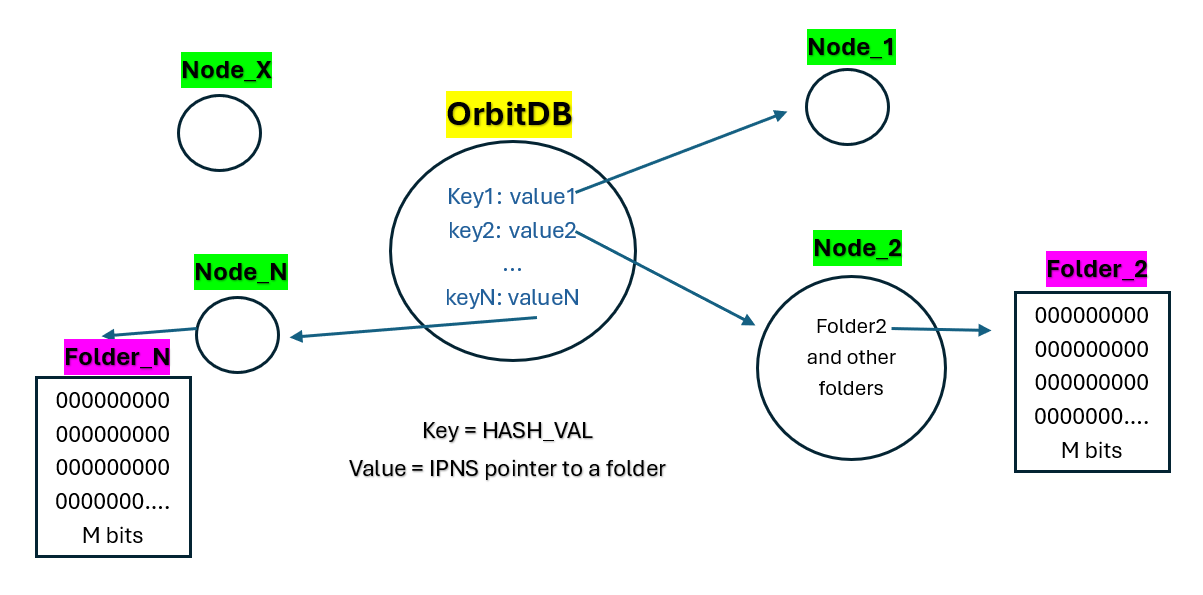}
    \caption{Step 1 - System Initialization}
    \label{fig:init}
\end{figure}

During initialization, the system creates a distributed database with \(N\) files, each containing \(M\) bits, initially set to \(0\). It also generates \(N\) unique hash addresses as keys, each pointing to a node storing a file in the distributed network. This design ensures balanced load distribution, resilience to failures, and consistent data accessibility. (Ref. Fig \ref{fig:init})

\subsubsection{Step 2: Inserting a New Key}
\begin{figure}[h!]
    \centering
    \includegraphics[width=1.0\textwidth]{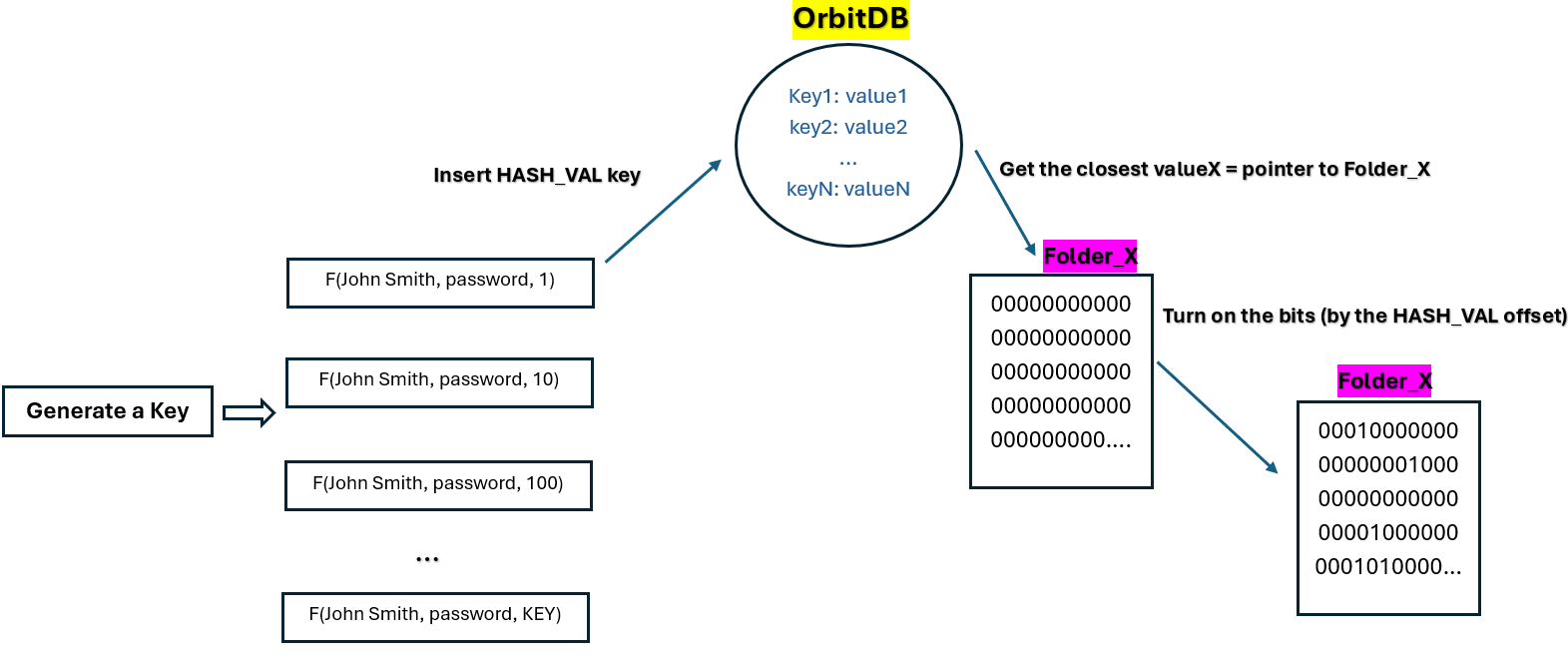}
    \caption{Step 2 - Insert a new key.}
    \label{fig:insert-operation}
\end{figure}

When a new key needs to be added to the system for a given username and password, the process follows these steps:

\begin{enumerate}
    \item \textbf{Generate a Random Key:}  
    The system generates a random key of length \(L\) bits (e.g., 16 bits).  
    For instance, suppose the generated key is \( \text{Key} = 1000111000010010 \).
    
    \item \textbf{Calculate Hash Values Using BFLUT:}  
    The system calculates Hash values for each prefix of the key, combining the username, password, and the prefix.  \\
    For example, for this \( \text{key} = 1000111000010010 \): 
\textit{    \( \text{F(John Smith, password, 1)} \)  
    \( \text{F(John Smith, password, 10)} \)   
    \( \text{F(John Smith, password, 100)} \) and etc.}
    
\item \textbf{Locate the Nearest File:}  
For each Hash value, the system uses OrbitDB to locate the nearest key by comparing the numerical distance between the target hash and the stored keys in the database. The ``nearest key'' is determined as the one with the smallest absolute difference from the target hash. This key serves as a pointer to the IPNS address of the file where data related to that prefix is stored. This ensures efficient lookup and retrieval of data distributed across the network.
    
    \item \textbf{Turn on the bits in the File:}  
    The system accesses the file via the IPNS address associated with the nearest key.  
    The Hash value is divided into equal parts, with each part representing locations in the file where bits are turned on.
    
    \item \textbf{Update the File:}  
    After turning on the bits, the file is saved back into the IPFS system, generating a new CID.  
    The IPNS address is updated to always point to the latest version of the file.  
    These steps repeat for all prefixes of the key until the insertion process is complete. (Ref. Fig \ref{fig:insert-operation})
\end{enumerate}

\subsubsection{Step 3: Retrieving a Key Using BFLUT}
\vspace{0\baselineskip} 
When a key needs to be retrieved from the system based on a username and password, the process follows these steps:

\begin{enumerate}
    \item \textbf{Calculate Hash Values for Retrieval:}  
    For the given username and password, the system calculates Hash values for all possible prefixes of the key.  
    For example:  
    \( \text{F(John Smith, password, 0)} \)  
    \( \text{F(John Smith, password, 1)} \)
    
    \item \textbf{Locate the Nearest File:}  
    For each Hash value, the system uses OrbitDB to locate the nearest key, which points to the relevant file where the bits are stored.
    
    \item \textbf{Verify the Presence of Information:}  
    The system accesses the file via the IPNS address and checks if the required bits are turned on at the appropriate locations.  
    If the bits are present, the information is considered valid, and the system proceeds to check the remaining prefixes.
    
    \item \textbf{Reconstruct the Key:}  
    The process continues for all prefixes until the full key of length \(L\) bits is reconstructed.  
    For example, if the bits for \( \text{PREFIX} = 1, 10, 100, 1000 \) are turned on and the required key length is 16 bits (and all prefixes are valid up to the 16th bit), the system successfully concludes that the entire key is present and retrieves the key \( \text{1000111000010010} \).  
\end{enumerate}

\subsection{Complete Example}

The following is a complete example of how the system works to insert a key and securely store its data across the network.

1. \textbf{Generating a New Key:}  
The system begins by generating a random Hash key and receiving the username and password as input. For example: \\ 
Username: \texttt{user123}, Password: \texttt{password123}, and Random Key: \texttt{ab12cd34ef56}.  
This unique key will be used to organize and locate information in the system.

2. \textbf{Hashing Prefixes:}  
The system iterates through each prefix of the generated key, combining the prefix with the username and password to calculate a Hash value. For the key \texttt{ab12cd34ef56}, the prefixes are \texttt{a}, \texttt{ab}, \texttt{ab1}, \texttt{ab12}, and so on up to the full key. For each prefix, the system computes a Hash using the formula:  
\[
\text{Hash} = \text{hash}(\text{username} + \text{password} + \text{prefix})
\]  
For example, for the prefix \texttt{ab12}, the computed Hash would be \texttt{5678abcd}.

3. \textbf{Locating the Nearest File:}  
For each calculated Hash, the system identifies the closest key in the database using Hash proximity. For example, if the Hash for the prefix \texttt{ab12} is \texttt{5678abcd}, and the database contains keys such as \texttt{1234abcd}, \texttt{5678efgh}, and \texttt{9101ijkl}, the closest key is determined to be \texttt{5678efgh}. The value associated with this key is the IPNS address of a file, which the system retrieves from the database.

4. \textbf{Updating the File:}  
The system accesses the file corresponding to the retrieved IPNS address and updates specific bits in the file based on the positions derived from the Hash values. For example, if the bit positions are \texttt{4, 8, 16}, these bits in the file are turned on. The updated file is uploaded back to IPFS, generating a new CID. The IPNS address for the key is updated to point to this new CID. This process continues for every prefix of the key, ensuring that all associated files are updated with consistent bit modifications based on the Hash values.

This iterative approach ensures that data is stored securely and that the system can later verify the information by recalculating and checking the specific bit patterns. By using the closest key mechanism, the system effectively distributes and organizes information across the network.

\subsection{Decentralized Write Operations Without OrbitDB}

OrbitDB serves as a platform for key storage in our decentralized system. It can be replaced with a more distributed approach, where each node maintains a table mapping addresses to specific nodes, ensuring efficient data distribution.

Inspired by Ripple’s trust model, we propose publishing a list of trusted nodes in a public and verifiable manner, akin to a blockchain. This approach ensures transparency and trust within the decentralized system.

When a user needs to write data, they first compute the relevant target nodes responsible for storing the bits. The user then directly interacts with these nodes to store the data, which improves performance.

This method prevents a single point of control over stored information, ensuring that data remains truly decentralized. Furthermore, a protocol can be defined to regulate write permissions, controlling the frequency and conditions under which data can be written. By implementing such mechanisms, we establish a fair, scalable, and secure decentralized storage system.

\section{Analysis}

In this study, we analyze a system based on the BFLUT data structure, designed to securely and efficiently store keys in a decentralized manner. The system operates by activating bits in files using unique keys and a specific Hash process. However, due to the characteristics of the data structure, there is a probability of encountering a False Positive where a key may point to data that does not actually belong to the intended user.

This section aims to analyze the probability of False Positives under different scenarios, identify the system's limitations, and optimize various parameters. Additionally, the analysis aims to evaluate the number of file accesses required to extract a single key based on the given system parameters. \\

For the purpose of this analysis, we make the following assumptions:
\begin{enumerate}
    \item All files in the system are treated as a single unified file, with a length equal to the sum of all individual file lengths.
    \item The processes of bit activation and retrieval are applied uniformly across the unified file.
\end{enumerate}

\subsection{Analysis Overview}

Before delving into the specific analyses, we define the key parameters used throughout this research:

\begin{itemize}
    \item \(N\): The total number of keys stored currently in the system.
    \item \(F\): The total number of bits in the system, calculated as the product of the number of files and the length of each file.
    \item \(L\): The length of each key, typically represented in binary.
    \item \(U\): The number of segments into which the Hash address is divided, used to determine the bit activation for each prefix.
\end{itemize}

These analyses provide insights into balancing security, scalability, and resource efficiency in decentralized environments.

\subsection{General Calculations}

1. For a single key, in one iteration of \(PREFIX\), the number of bits activated is:
   \[
   \frac{L}{U}
   \]

2. Over all iterations across the entire key (\(L\)), the total number of bits activated is:
   \[
   L \cdot \frac{L}{U}
   \]

3. The probability that a specific bit remains unactivated by a single key is:
   \[
   1 - \frac{L \cdot \frac{L}{U}}{F}
   \]

4. The probability that a specific bit remains unactivated after \(N\) keys is:
   \[
   \left(1 - \frac{L \cdot \frac{L}{U}}{F}\right)^N
   \]

5. The probability that any bit is activated at least once:
   \[
   1 - \left(1 - \frac{L \cdot \frac{L}{U}}{F}\right)^N
   \]

6. The probability that all the bits activated by the next key (\(N+1\)) are already activated:
   \[
   \left(1 - \left(1 - \frac{L \cdot \frac{L}{U}}{F}\right)^N\right)^{L \cdot \frac{L}{U}}
   \]
This represents the probability of a False Positive. \\
Specifically, for a search of length \(L\), if in every iteration of the search all the bits corresponding to the key's prefixes are lit in their designated positions (as calculated in step 6), the system will incorrectly determine that the key exists. This happens due to overlaps caused by other inserted keys activating those specific bits, leading to a False Positive result.\\
The probability of this happening depends on the number of keys already inserted (\(N\)), the total memory size (\(F\)), and the ratio of activated bits (\(L/U\)), as described in the formula for step 6.

\subsection{Analysis Details}
In the following analysis, the keys will be treated as having a length of 64 hexadecimal characters (equivalent to 256 bits).
Each analysis will focus on specific parameters by fixing some variables and substituting relevant values:

\subsubsection{Analysis 1: Balancing Key Length (\(L\)) and Segment Division (\(U\))}

When analyzing the ratio of activated bits (\(\alpha\)) in distributed systems utilizing the BFLUT mechanism, the selection of parameters \(U\) and \(L\) significantly impacts system behavior. Notably, the base of the key representation (binary, hexadecimal, etc.) does not directly affect system performance, as the activated bits are treated as indices. Thus, the focus is placed on the number of activated bits and their ratio relative to the system parameters.

The activated bit ratio (\(\alpha\)) represents the proportion of bits in the system that are turned on, serving as a critical parameter for balancing system efficiency and the probability of False Positives (\(P_{FP}\)). A commonly desired ratio is \(\alpha = 0.5\), where half the bits in the system are activated. This ratio helps reduce the probability of False Positives while maintaining sufficient inactive bits to support efficient searches.

To achieve the desired \(\alpha\), the number of activated bits is calculated as:

\[
\alpha = \frac{\text{Activated Bits}}{F} = \frac{N \cdot L^2 \cdot \frac{1}{U}}{F},
\]

where \(N\) is the number of keys, \(L\) is the key length, \(U\) is the number of segments, and \(F\) is the total number of bits in the system. Solving for \(U\):

\[
U = \frac{N \cdot L^2}{\alpha \cdot F}.
\]
A higher \(U\) value reduces activated bits per iteration, lowering False Positives but also decreasing validations, which may cause early failures. Conversely, a lower \(U\) value increases activations, improving validation but raising False Positives.

Similarly, increasing \(L\) raises the False Positive rate, emphasizing the need to optimize \(U\) based on system-specific factors like the number of keys (\(N\)), memory size (\(F\)), and efficiency requirements.

\subsubsection{Analysis 2: Effect of Number of Keys \(N\)}
We will analyze the impact of \(N\) on the probability of False Positives while keeping \(L = 64\), \(U = 4\), and \(F = 2^{21} \cdot 150\) (representing the size of each file in bits multiplied by the number of files).

Using these values, the formula for the probability of False Positives becomes:

\[
\left(1 - \left(1 - \frac{64 \cdot \frac{64}{4}}{2^{21} \cdot 150}\right)^N\right)^{64 \cdot \frac{64}{4}}
\]

As \(N\) increases, the probability of a False Positive also increases, as more keys activate additional bits in the system, raising the chance of collisions. Below are the probabilities for different values of \(N\):

\begin{itemize}
    \item For example \(N = 500,000\): The probability of a False Positive is approximately \(5.77 \times 10^{-93}\) (negligible).

\end{itemize}

\begin{table}[h]
    \centering
    \begin{tabular}{|c|c|}
        \hline
        \textbf{N} & \textbf{Computed Result} \\ 
        \hline
        100000  & 0.00e+00 \\
        200000  & 0.00e+00 \\
        300000  & 6.91e-211 \\
        400000  & 6.99e-142 \\
        500000  & 5.77e-98 \\
        600000  & 9.53e-69 \\
        700000  & 8.83e-49 \\
        800000  & 6.71e-35 \\
        900000  & 3.87e-25 \\
        1000000 & 3.21e-18 \\
        \hline
    \end{tabular}
    \vspace{2mm}
    \caption{False Positive Probability as a Function of \(N\).}
    \label{tab:computed_results}
\end{table}

The table provides a detailed breakdown of the calculations for all the tested values of \(N\).

For a detailed visualization, refer to Appendix~\ref{sec:appendix}.

This demonstrates that as the number of keys grows, the system's ability to distinguish between keys diminishes, leading to a higher likelihood of False Positives.

\subsubsection{Analysis 3: Minimum Storage Size (\(F\)) for Given \(P_{FP}\)}

In this analysis, we calculate the minimum \(F\) required to support \(N\) keys with a specified False Positive probability (\(P_{FP}\)). By fixing the parameters \(N, L, U\), we evaluate the required storage size \(F\) for varying \(P_{FP}\) values.

The formula to calculate \(F\) is derived as follows:

\[
F \geq \frac{L^2}{U \cdot \left(1 - \left(1 - P_{FP}^{\frac{U}{L^2}}\right)^{\frac{1}{N}}\right)}
\]

We fixed \(N = 500,000\), \(L = 64\), and \(U = 4\) for this calculation. The required \(F\) values for different False Positive probabilities (\(P_{FP}\)) are shown below:
\begin{itemize}
\item For \(P_{FP} = 10^{-6}\): The required \(F\) is approximately \(56.61 \cdot 2^{21}\) bytes.
\item For \(P_{FP} = 10^{-9}\): The required \(F\) is approximately \(62.43 \cdot 2^{21}\) bytes.
\item For \(P_{FP} = 10^{-12}\): The required \(F\) is approximately \(67.33 \cdot 2^{21}\) bytes.
\end{itemize}

These results demonstrate the relationship between the desired false-positive probability and the required storage size (\(F\)). Each file in the system is assumed to have a size of \(2^{21}\) bytes. Thus, the system will require between 55 and 70 files for the given probabilities. As \(P_{FP}\) decreases (i.e., stricter False Positive requirements), the required storage size increases significantly.
For a detailed visualization, refer to Appendix~\ref{sec:appendix}.

\subsubsection{Analysis 4: Analysis of Expected Number of File Accesses}
In a distributed system, the expected number of file accesses is a critical metric for understanding the efficiency and resource usage of the system. This metric helps evaluate the system's performance and scalability. \\
Formally, for a discrete random variable \(X\), the expected value is defined as:\\
\[
E[X] = \sum_{x \in \text{Range}(X)} x \cdot P(X = x)
\]
where \(P(X = x)\) is the probability that \(X\) takes the value \(x\).

\paragraph{Problem Statement}
Suppose the system contains \(K\) files in total and a key length  \(L\). These accesses may include repeated visits to the same file. The goal is to calculate the expected number of unique files accessed when retrieving data for a key.

\paragraph{Derivation of the Expected Number of Unique Accesses}

In this calculation, we consider the key length to be 256 bits.  
At each step, we check both possible values (0 and 1), meaning each step involves two operations.  
Since the total number of steps is equal to the key length, the total number of operations is 2L = 256 * 2

Thus, the expected number of unique files when we have a system with 16 files is given by:

\[
E[X] = K \cdot \left(1 - \left(1 - \frac{1}{K}\right)^{2L}\right)
\]

If we also consider false positives, there may be a few additional file accesses.  
However, if \( P_{FP} \) is very small, this additional number of accesses remains negligible.

\subsection{Simulation}

In this simulation, we examined the process of key extraction in a distributed system using IPFS, IPNS, and OrbitDB. The goal was to reconstruct a key while minimizing file accesses.

\subsubsection{Key Findings}
Initially, in our simulation, we set \( m \), the number of files, to 100.
Unlike binary search (logarithmic complexity, checking two options per step), our approach expanded the search space by testing all 16 possible hexadecimal characters (0 to f) at each step. This significantly increased the initial search size.

Each extraction attempt validated all 16 extensions, causing an exponential growth in checks. Unlike binary search, where each step halves the search space, our method evaluated all potential values, leading to a higher number of unique file accesses.\\

\begin{table}[h]
    \centering
    \begin{tabular}{|c|c|c|}
        \hline
        \textbf{Username} & \textbf{Password} & \textbf{NumberOfUniqueFiles} \\
        \hline
        ``alice12''       & securePass1!    & 95 \\
        bob\_smith      & bobRocks42@     & 98 \\
        charlie.dev    & charlieCode99\$ & 96 \\
        david\_w        & DavidPass123*   & 97 \\
        emma.l         & emmaLovesCats!  & 95 \\
        frank\_t        & FrankStrongP@ss & 97 \\
        grace.hopper   & graceCode42\#   & 95 \\
        henry\_m        & HenrySafePass1! & 96 \\
        isabella\_99    & BellaSecret\$22 & 98 \\
        jack\_admin     & AdminJack\#2024 & 94 \\
        \hline
    \end{tabular}
        \vspace{3mm}
        \caption{Simulation results per user (unique file accesses).}
    \label{tab:user_data}
\end{table}

Realizing that the number of file accesses was likely to be very high, especially given the large number of files, we adjusted our approach. We reduced the number of files to 50 while increasing their size. This will limit the number of accessed files to 50.

Since multiple files may reside on the same cluster (node), the actual number of physical accesses is much lower. Finding the right balance between the number of files and their size is key to optimizing performance and secure distribution.

\section{Discussion and Conclusions}
\label{s:dc}

\textbf{Secure Key Management Without Centralized Storage}  
The proposed system effectively manages cryptographic keys in a decentralized environment without explicitly storing them. Leveraging BFLUT and IPFS ensures that key recovery remains secure even if multiple or even all nodes are compromised, as long as the prefix used in the hash, for example, is the username, and the password is unknown. The bits maintained by each node result from a cryptographic hash function (SHA) and thus do not leak information. Furthermore, when nodes may act adversarially, information retrieval is still possible by adding intermediate error-correcting codes (during the writes), correcting partial retrievals, and continuing to retrieve the entire information. Denial of service regarding too many writes by (non-allowed) users should be monitored and restricted by each node to ensure a limited number of writes. As for reads, the distribution of nodes that maintain the database and the possibility to tolerate errors yields a possibility to cope with erasures, too. More details are deferred to the full version.

\noindent
\textbf{Optimizing File Count and Size for Performance}  
Initial simulations with 100 files resulted in excessive file accesses. By reducing the number of files to 50 while increasing their size, the system achieved \textbf{47-50} accesses per run, striking a better balance between efficiency and resource utilization.

\appendix
\section{Appendix}
\label{sec:appendix}

\begin{figure}[H]
    \centering
    \includegraphics[width=1.0\textwidth]{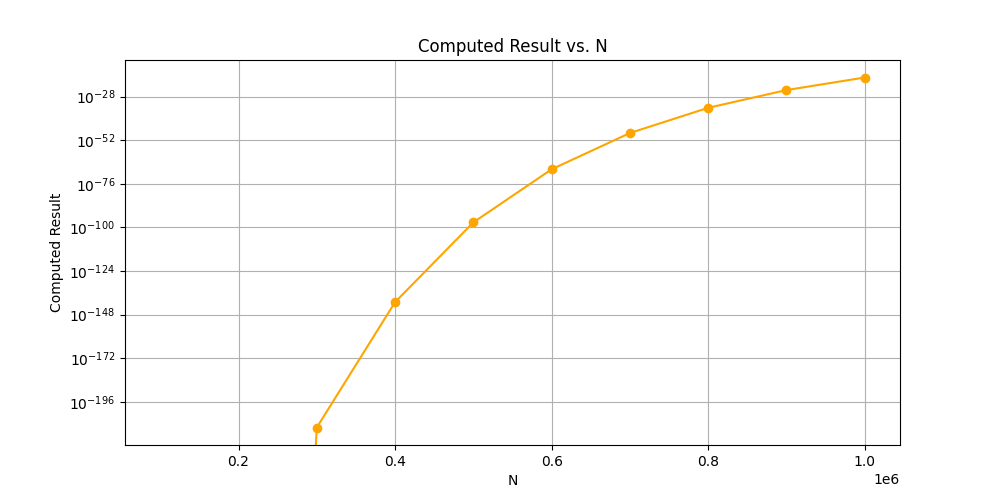}
    \caption{False Positive Probability as a Function of \(N\).}
    \label{fig:analyze1}
\end{figure}

\begin{figure}[H]
    \centering
    \includegraphics[width=1.0\linewidth]{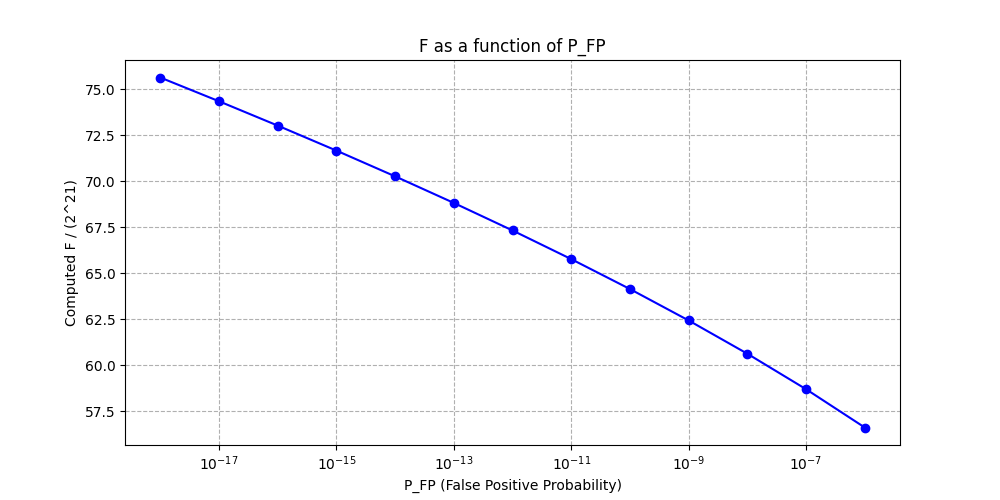}
    \caption{Required \(F\) as a function of \(P_{FP}\) for fixed \(N = 500,000\), \(L = 64\), and \(U = 4\).}
    \label{fig:appendix_analysis3}
\end{figure}

\clearpage
\bibliographystyle{unsrt}
\bibliography{arXivDGS}  

\end{document}